\documentclass{elsart}
\usepackage{graphicx,amsmath,amssymb}
\journal{Physica A}

\begin{document}
\begin{frontmatter}
\title{The 3-dimensional random walk with applications to overstretched DNA and the protein titin}
\author[erik]{Erik Van der Straeten}, \thanks[erik]{Research Assistant of the Research
Foundation - Flanders (FWO - Vlaanderen)}
\ead{Erik.VanderStraeten@ua.ac.be}
\author{Jan Naudts}\ead{Jan.Naudts@ua.ac.be}
\address{Departement Fysica, Universiteit Antwerpen, Groenenborgerlaan 171, 2020 Antwerpen, Belgium}

\begin{abstract}
We study the three-dimensional persistent random walk with drift. Then we develop a thermodynamic model that is based on this random walk without assuming the Boltzmann-Gibbs form for the equilibrium distribution. The simplicity of the model allows us to perform all calculations in closed form. We show that, despite its simplicity, the model can be used to describe different polymer stretching experiments. We study the reversible overstretching transition of DNA and the static force-extension relation of the protein titin.
\end{abstract}
\end{frontmatter}

\section{Introduction}
Nowadays, one can experimentally manipulate individual molecules. One can measure force-extension relations of biopolymers under varying circumstances. Double stranded DNA is intensively studied in the literature. Mostly the Worm Like Chain model is used to describe the properties of DNA in a good solvent. The theoretical and experimental force-extension relations are in good agreement at relatively low forces \cite{wlc_fit}. When one applies a large force (around $65$ pN) on double stranded DNA, one observes an abrupt increase of the contour length of the molecule \cite{exp_cluzel,exp_bustam}. This marks a transition from the standard B-form of DNA to a so called S-form of DNA. In \cite{exp_bustam}, one performs stretching and releasing experiments on the same molecule under different solvent conditions. Often the stretch and release curves do not coincide and hysteresis is observed. One can eliminate this hysteresis by increasing the salt concentration of the solvent. Then one can assume that the experiment is performed under equilibrium conditions. The salt dependence of the overstretching transition is further studied in \cite{exp_bloom}. The temperature dependence of this transition is studied in \cite{exp_bloom2}.

The reversible B-DNA to S-DNA transition is studied theoretically by different authors. Typically the developed models contain several adjustable parameters. We will make a strict distinction between 'a posteriori' (called posteriors) and 'a priori' parameters (called priors). The former have the usual meaning of parameters that are changed in order to obtain a good fit of a theoretical curve to an experimental dataset. The latter have the meaning that the values of the parameters can be considered as fixed and are not used in the fitting procedure.  In an attempt to explain the obtained experimental data of the B-DNA to S-DNA transition, a pure two-state model is used in \cite{exp_cluzel}. The application of this model is limited to the transition region. For this reason, the model is combined in \cite{theo_bruins} with the well known Worm Like Chain model. The combination of the two models results in a good fit to the experimental data of \cite{exp_cluzel} with $2$ posteriors and $3$ priors. Another theoretical model is introduced in \cite{theo_storm}. This so called Discrete Persistent Chain model, combines features from the Worm Like Chain model and the Freely Jointed Chain model and contains $7$ posteriors (although some of them could have been used as priors). As a consequence of the large number of parameters, it is no surprise that a very good fit to one of the experimental datasets of \cite{exp_bustam} is obtained. In \cite{theo_punk}, the authors argue that one has to include salt effects to explain all the datasets of \cite{exp_bloom}, obtained under different solvent conditions. Starting from a phenomenological expression for the free energy, the authors obtain theoretical curves that depend on $4$ posteriors and several priors. A good fit to $7$ different datasets of \cite{exp_bloom} is obtained. The fit parameter that is the most sensitive to the salt concentration is the effective length of charge separation. The range of values of this parameter obtained by the fit to the $7$ datasets is in agreement with previous studies. The authors conclude that their ansatz for the free energy is physically meaningful.

Another molecule that is often studied in the literature is the protein titin. It contains approximately $30.000$ amino acids. The most important part of this protein is the so called PEVK region that is flanked by immunoglobulin domains. The PEVK region is a chain of amino acids that behaves like a random coil. The immunoglobulin domains are folded parts of the protein. In \cite{tsk}, the static (equilibrium) force-extension relation is measured. Although the resulting curve is non-trivial, surprisingly one usually only studies in the literature \cite{tit_rev} the dynamic force-extension relation of this protein.
In this paper we will develop a thermodynamic model that is based on the three-dimensional random walk,
to describe the static force-extension relation of titin.

In \cite{erik_atwo}, the present authors obtained analytical results for the one-dimen\-si\-onal persistent random walk with drift. It is shown in \cite{erik_aone,erik_resi} that this walk can be used as a qualitative model for a polymer in solution. In the present paper we study the three-dimensional persistent random walk with drift. We will show that it can be used to study the force-extension relations of overstretched DNA and of the protein titin. The persistent random walk is intensively studied in the literature \cite{per1,per2,per3,per4}. However, to the best of our knowledge, this is the first time that the persistent random walk is used to study the aforementioned force-extension relations.

In the next section our mathematical model is introduced. The section starts with a brief summary of the most important results of \cite{erik_tran}. The application of these results to the three-dimensional persistent random walk with drift then follows. In section \ref{the_mod}, we establish the connection between the parameters of our mathematical model and the thermodynamic control parameters of interest. In sections \ref{SB_overgang} and \ref{appl_tit}, the applications of our formalism are studied. The last section gives a short discussion of the results.

\section{Mathematical model}
In \cite{erik_tran} Markov chains with a finite number of states are studied. A Markov chain with state space $\Gamma$ is determined by initial probabilities $p(x)$ ($x\in\Gamma$), and by transition probabilities $w(x,y)$ ($x,y\in\Gamma$), with
\begin{eqnarray}\label{normpw}
1=\sum_{x\in\Gamma}p(x)&\textrm{and}&1=\sum_{y\in\Gamma}w(x,y).
\end{eqnarray}
The probabilities $p(x)$ are used as initial values for the equation of motion
\begin{eqnarray}
p_{t+1}(x)&=&\sum_{y\in\Gamma}p_t(y)w(y,x).
\end{eqnarray}
They are called stationary if the following equation holds
\begin{eqnarray}\label{stat_prob}
p(x)&=&\sum_{y\in\Gamma}p(y)w(y,x),
\end{eqnarray}
The record of transitions $k$ is defined \cite{erik_tran} as a sequence of numbers $k_{x,y}$, one for each pair of states $x,y$, counting how many times the transition from $x$ to $y$ is contained in a given path of the Markov chain. One can prove following properties for stationary Markov chains \cite{erik_tran}
\begin{eqnarray}\label{eig_random}
\langle k_{x,y}\rangle&=&np(x)w(x,y)
\cr 
S&=&-n\sum_{x,y\in\Gamma}p(x)w(x,y)\ln w(x,y)
\cr
&&-\sum_{x\in\Gamma}p(x)\ln p(x),
\end{eqnarray}
with $n$ the total number of chain elements, $S$ the standard Boltzmann-Gibbs entropy and $\langle .\rangle$ an average over phase space.

Consider now a discrete, three-dimensional random walk with transition probabilities which depend only on the direction of the present and of the previous step. This means that the process of the increments is Markovian. The state space of the latter process contains $6$ elements and is
\begin{eqnarray}
\Gamma&=&\{x+,x-,y+,y-,z+,z-\}.
\end{eqnarray}
To clarify this notation, $x+$ has the meaning of a step in the positive $x$-direction, while $x-$ is a step in the negative $x$-direction. The Markov chain is determined by $6\times6=36$ different transition probabilities. To reduce this number we first assume that the walker cannot turn back. This is equivalent with following constraints on the transition probabilities
\begin{eqnarray}
0&=&w(x+,x-)=w(y+,y-)=w(z+,z-)
\cr
&=&w(x-,x+)=w(y-,y+)=w(z-,z+).
\end{eqnarray}
For the applications, studied in the present paper, it is important to break the symmetry in one of the spatial directions {\it and} to distinguish between steps that go straight on or change direction. So, to further reduce the number of transition probabilities we can assume that the states $y+$, $y-$, $z+$ and $z-$ are equivalent. Then define the following shorthand notation
\begin{eqnarray}\label{syymmm}
\epsilon&:=&w(x+,y+)=w(x+,y-)=w(x+,z+)=w(x+,z-)
\cr
\mu&:=&w(x-,y+)=w(x-,y-)=w(x-,z+)=w(x-,z-)
\cr 
\alpha&:=&w(y+,x+)=w(y-,x+)=w(z+,x+)=w(z-,x+)
\cr 
\gamma&:=&w(y+,x-)=w(y-,x-)=w(z+,x-)=w(z-,x-)
\cr 
\theta&:=&w(y+,z+)=w(y-,z+)=w(y+,z-)=w(y-,z-)
\cr 
&=&w(z+,y+)=w(z-,y+)=w(z+,y-)=w(z-,y-).
\end{eqnarray}
\begin{figure*}
\begin{center}
\parbox{0.49\textwidth}{\includegraphics[width=0.48\textwidth]{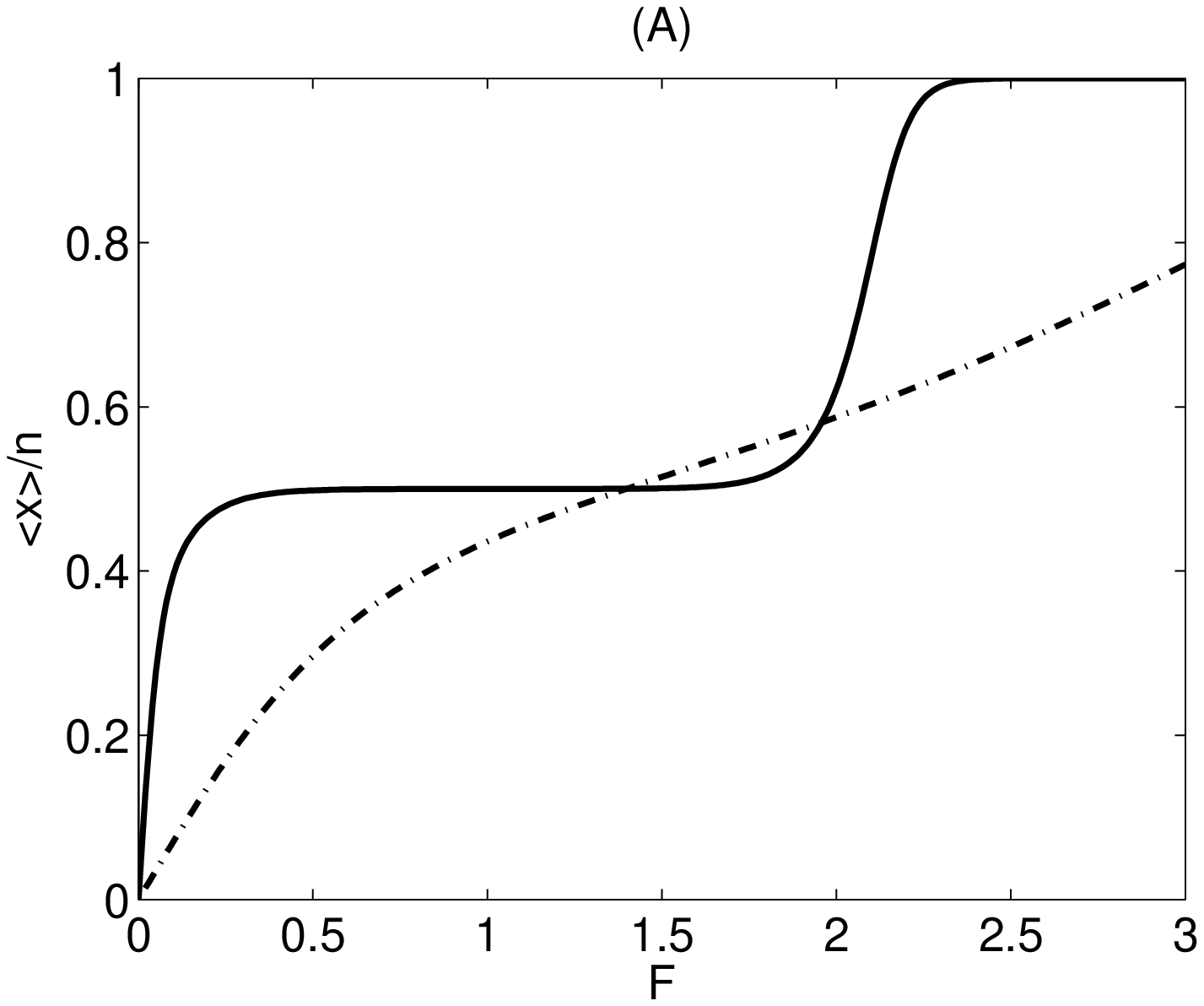}}
\hfill
\parbox{0.49\textwidth}{\includegraphics[width=0.48\textwidth]{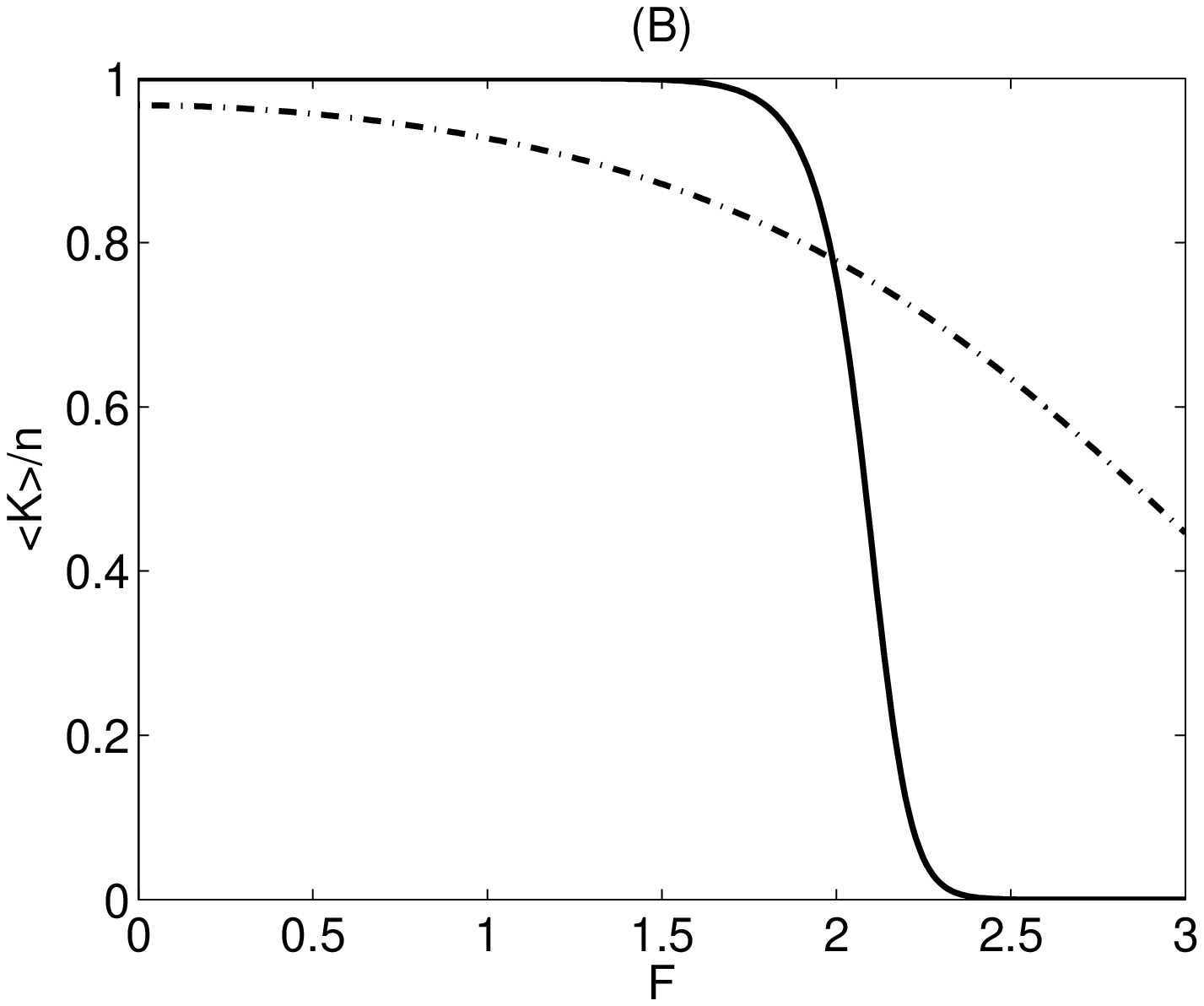}}
\hfill
\parbox{0.49\textwidth}{\includegraphics[width=0.48\textwidth]{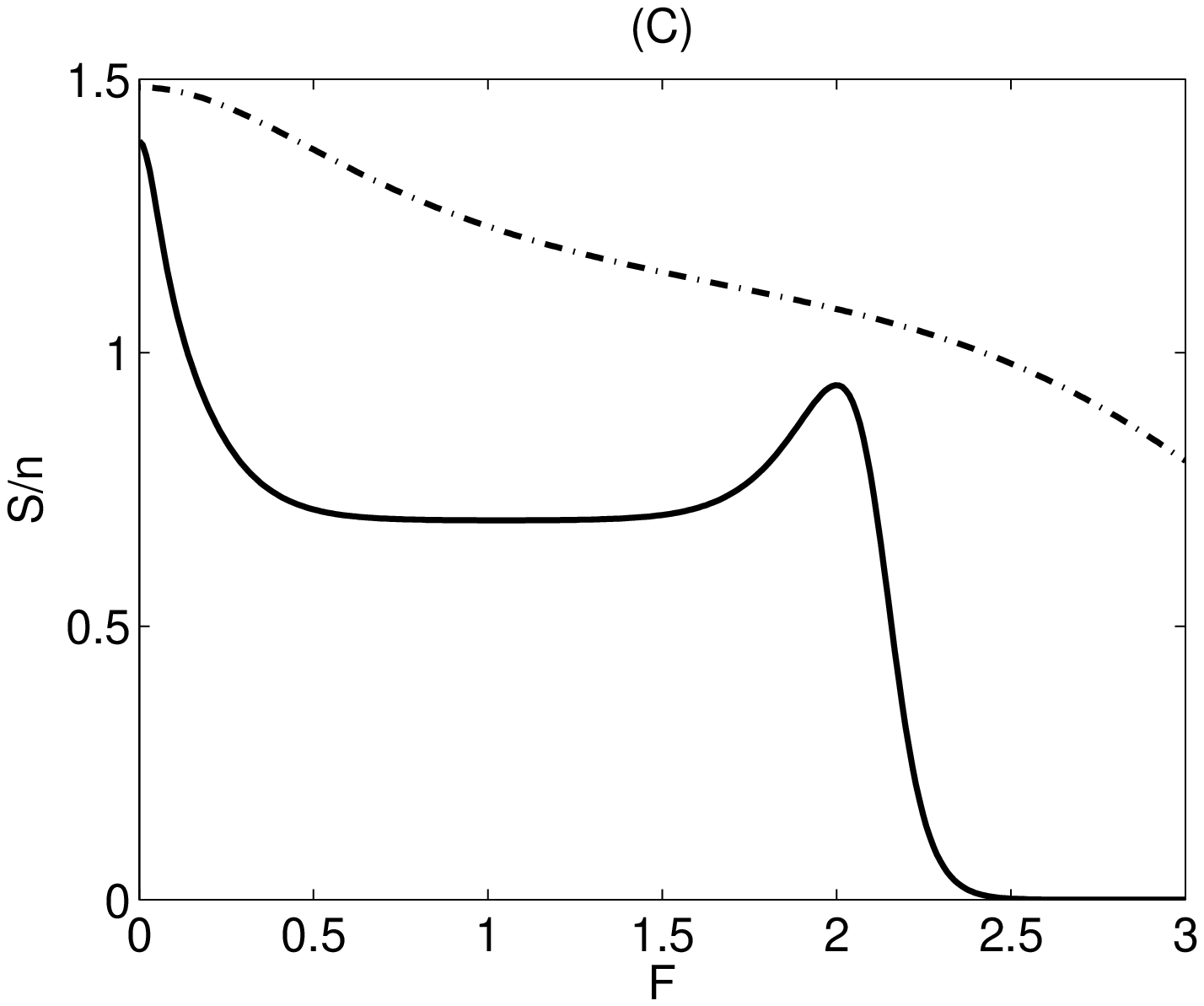}}
\caption{\label{fig:exte}Plot of the average endposition (A), the average number of changes of direction (B) and the entropy (C) as a function of the force. For all figures, the value of the temperature equals  $T=0.05$ for the solid line and $T=0.5$ for the dotted line. ($h=-1$, $a=b=1$)} 
\end{center}
\end{figure*}Using the normalisation condition (\ref{normpw}) one obtains following expressions for the remaining $6$ transition probabilities
\begin{eqnarray}
&&w(y+,y+)=w(z+,z+)=w(y-,y-)=w(z-,z-)=1-\alpha-\gamma-2\theta
\cr 
&&w(x+,x+)=1-4\epsilon,\ \ w(x-,x-)=1-4\mu.
\end{eqnarray}
So now we are left with a $5$-parameter model. The stationary probabilities can be calculated by solving the set of equations (\ref{stat_prob}) together with the normalisation condition (\ref{normpw}). The result is
\begin{eqnarray}
&&p(x+)=\frac1N\alpha\mu,\ \ p(x-)=\frac1N\gamma\epsilon,
\cr
&&p(y\pm)=p(z\pm)=\frac1N\epsilon\mu,
\end{eqnarray}
with $N=4\epsilon\mu+\alpha\mu+\gamma\epsilon$. Then the average number of changes of direction $\langle K\rangle$ and the entropy $S$ can be calculated with (\ref{eig_random})
\begin{eqnarray}\label{KS}
\frac{\langle K\rangle}{n}&=&1-\frac1n\sum_{i\in\Gamma}\langle k_{i,i}\rangle=8\epsilon\mu\frac1N\left(\alpha+\gamma+\theta\right),
\cr 
\frac Sn&=&-p(x+)\left[(1-4\epsilon)\ln(1-4\epsilon)+4\epsilon\ln\epsilon\right]
\cr
&&-p(x-)\left[(1-4\mu)\ln(1-4\mu)+4\mu\ln\mu\right]
\cr 
&&-4p(y+)\big[(1-\alpha-\gamma-2\theta)\ln(1-\alpha-\gamma-2\theta)
\cr
&&+\alpha\ln\alpha+\gamma\ln\gamma+2\theta\ln\theta\big],
\end{eqnarray}
where finite size corrections are ignored in the expression for the entropy. We now introduce an extra asymmetry in the model. We use two different lattice parameters $a$ en $b$. If the walker goes straight on, the length of the step is $a$, if the walker changes direction, the length of the step is $b$. The $x$-component of the endposition of the walk is
\begin{eqnarray}
x&=&a(k_{x+,x+}-k_{x-,x-})+b\sum_{i\in\Gamma'}\left(k_{i,x+}-k_{i,x-}\right)
\end{eqnarray}
with $\Gamma'=\Gamma\setminus\{x+,x-\}$. The average of $x$ becomes
\begin{eqnarray}\label{x}
\frac{\langle x\rangle}{n}&=&\frac aN\left[\alpha\mu(1-4\epsilon)-\gamma\epsilon(1-4\mu)\right]+\frac bN4\epsilon\mu(\alpha-\gamma).
\end{eqnarray}
The averages of the $y$- and $z$-component of the endpostion of the walk vanish because of the imposed symmetry.

\section{Thermodynamic model}\label{the_mod}
In this section, we follow the lines of \cite{erik_resi} to obtain expressions for the thermodynamic control parameters of interest as a function of $\epsilon,\ \mu,\ \alpha,\ \gamma,\ \theta$. In experiments one measures force-extension relations of biopolymers at constant temperature. To establish the connection between the random walk and the polymer, we interpret the endposition of the walk as the extension of the molecule. The corresponding control parameter is the force $F$ applied to the endpoint of the molecule. The Hamiltonian is defined as $H=hK$, with $K$ the number of changes of direction and $h$ a constant with dimensions of energy. Depending on the sign of $h$, the ground state of the system is a compact walk that always changes direction ($h<0$) or a stretched walk that always goes straight on ($h>0$). The reason why we define the Hamiltonian like this will become clear in the following sections. The control parameter corresponding with the energy $E=h\langle K\rangle$, is the temperature $T$. The Legendre transform of the entropy $S$ is the free energy $G$
\begin{eqnarray}
G&=&\min_{\epsilon,\mu,\alpha,\gamma,\theta}\left\{E-F\langle
x\rangle-\frac{1}{\beta}S\right\}.
\end{eqnarray}
Explicit expressions for $E=h\langle K\rangle$, $\langle x\rangle$ and $S$ as a function of the model parameters are known, see (\ref{KS}) and (\ref{x}). As a consequence, the solution of the set of equations
\begin{eqnarray}
\frac{\partial G}{\partial\epsilon}=0,\ \frac{\partial G}{\partial\mu}=0,\ \frac{\partial G}{\partial\alpha}=0,\ \frac{\partial G}{\partial\gamma}=0\ \textrm{and}\ \frac{\partial G}{\partial\theta}=0,
\end{eqnarray}
can be obtained in closed form
\begin{eqnarray}\label{minim}
\beta[-h+F(b-a)]&=&\ln\frac{\epsilon}{1-4\epsilon}\frac \alpha\theta
\cr
h\beta&=&\ln\frac{1-\alpha-\gamma-2\theta}\theta,
\cr
2\beta Fb&=&\ln\frac\epsilon\mu\frac\alpha\gamma
\cr
2\beta Fa&=&\ln\frac{1-4\epsilon}{1-4\mu},
\cr
(1-\alpha-\gamma-2\theta)^2&=&(1-4\epsilon)(1-4\mu).
\end{eqnarray}
This set of equations can be inverted and has a unique physical solution for every value of $\beta$ and $F$, see appendix \ref{app}. Then a plot of the average endposition of the walk as a function of the external force at constant temperature can be obtained, see figure \ref{fig:exte}A. When $h<0$, the average endpostion increases in two steps at low temperatures. At high temperatures this multi-step behaviour disappears. It is useful to study the average number of changes of direction and the entropy as a function of the force to understand this non-trivial behaviour, see figures \ref{fig:exte}B and \ref{fig:exte}C.

The steep increase of the average endposition at low forces is a result of the degeneracy of the ground state of the model. Contrary to the one-dimensional case, the three-dimensional random walk has different compact configurations for which the number of changes of direction equals the total number of steps. At vanishing force, the system has no preference for any of these configurations which results in a vanishing average endposition and a high value of the entropy. At low force, the walk can lower its free energy by choosing the configuration with the largest extension without changing the number of changes of direction. As a consequence one observes an increase of $\langle x\rangle$, a decrease of $S$, while $\langle K\rangle$ remains approximately constant.

\begin{figure}
\begin{center}
\parbox{0.49\textwidth}{\includegraphics[width=0.48\textwidth]{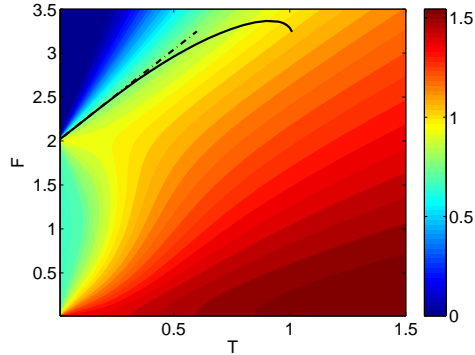}}
\caption{\label{fig:phase}Plot of the entropy as a function of the temperature and the force. The colour code is mentioned to the right. The black solid line, marks the gradual transition from the compact phase to the stretched phases. The black dotted line is an approximation for the solid line, valid at low temperatures only. ($h=-1$, $a=b=1$)} 
\end{center}
\end{figure}

The steep increase of the average endposition at intermediate force is also present in the one-dimensional random walk \cite{erik_resi}. It is the result of the competition between the folding energy $E$ and the potential energy $-F\langle x\rangle$. In \cite{erik_resi}, this sudden change of the average endposition is used to define the gradual transition from compact to stretched state of the walk. The boundary line is obtained by calculating the peak value of $\partial\langle x\rangle/\partial F$ at constant temperature. For the present model, the same criterion is used. Figure \ref{fig:phase} shows the entropy as a function of force and temperature. The black solid line shows the boundary between the two phases and is obtained by a numerical calculation. An unexpected result is that the boundary line is an increasing function of the temperature at low temperatures. This is also observed in the one-dimensional persistent random walk \cite{erik_resi} and in self avoiding random walks \cite{saw_kumar,saw_maren} and is a consequence of a subtle asymmetry in the entropy that favours the compact phase above the stretched phase. This asymmetry can be seen in figures \ref{fig:exte}C and \ref{fig:phase}. At low temperatures, the entropy is clearly not symmetric around $F=2$.

\begin{figure}
\begin{center}
\parbox{0.49\textwidth}{\includegraphics[width=0.48\textwidth]{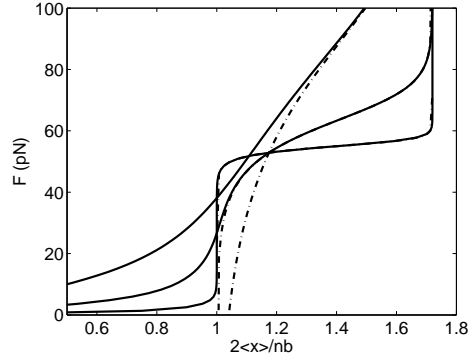}}
\caption{\label{fig:approx}The exact force-extension relation (solid lines) together with the approximate relation (dotted lines) obtained for different values of the parameter $-h\beta$ (top) 1;5;24.67 (bottom). For all curves following values of the remaining parameters are used $a=1.72$, $b=2$, $h=-19.04b$.} 
\end{center}
\end{figure}

In the appendix \ref{app2} an approximate expression for the boundary line at low temperatures is calculated, under the assumptions $h<0$, $a\leq b<2a$. One obtains $F=2+2T+\ldots$, with $h=-1$ and $a=b=1$. Figure \ref{fig:phase} shows  the result of the latter expression together with the exact boundary line. The two lines coincide up to a temperature of approximately $0.5$. An approximate formula for the force-extension relation in the transition region ($b/2<\langle x\rangle/n<a$) is also derived in appendix \ref{app2}. The result of the latter expression and the exact force-extension relation is shown in figure \ref{fig:approx} for different values of the parameter $h\beta$. Clearly the approximation becomes better for higher values of $-h\beta$.

\section{B-DNA to S-DNA transition}\label{SB_overgang}
When one applies a large force on double stranded DNA, one observes a transition from the standard B-form of DNA (with length per base pair of $0.34$ nm) to a so called S-form of DNA (with length per base pair of $0.58$ nm). Introduce $L_B$ and $L_S$, the contour length of B-DNA and S-DNA respectively. Then, divide the DNA chain in short segments that are in the B-form or S-form. Then define $N_s$ and $N_{bp}$ as the number of segments and the total number of base pairs of the DNA chain respectively. 

\begin{figure}
\begin{center}
\parbox{0.49\textwidth}{\includegraphics[width=0.48\textwidth]{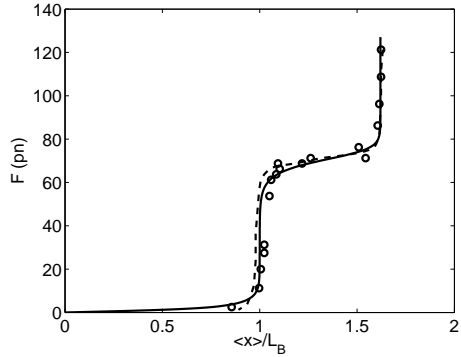}}
\caption{\label{fig:sb}Force-extension curves of a single double-stranded DNA molecule. ($\bullet\ \bullet\ \bullet$): Experimental data from \cite{exp_cluzel}. (solid, dotted line): Theoretical curves obtained with the present model and the model of \cite{theo_bruins} respectively.} 
\end{center}
\end{figure}

Now we use the three-dimensional walk as a model for the DNA chain. In our interpretation, changing direction during the walk corresponds with an B-segment of the DNA chain, while going straight on during the walk corresponds with an S-segment. Now it becomes clear why we defined the Hamiltonian as $H=hK$ with $K$ the number of changes of direction of the walk and $h$ a constant with dimensions of energy. When we chose the sign of $h$ negative, the ground state of the model is the desired standard B-form of DNA. We introduced two different lattice parameters $a$ and $b$. If the walker changes direction, the length of the step is $b$, if the walker goes straight on, the length of the step is $a$. As a consequence, the total number of steps of the walk $n$ and the number of segments of the DNA chain $N_s$ are equal. The contour length of the B-DNA and the S-DNA are connected to the lattice parameters of the random walk by
\begin{eqnarray}\label{connection_dna_walk}
L_B=\frac n2b=\frac{N_s}2b,&&L_S=na=N_sa.
\end{eqnarray}
In \cite{exp_cluzel}, the force-extension curve of a single DNA molecule is measured. An overstretching transition at $F\approx65$pN is observed. A theoretical model to describe this transition is introduced in \cite{theo_bruins}. This model contains $2$ posteriors and $3$ priors. The present model contains $3$ parameters $a$, $b$ and $h$. The ratio $a/b$ is considered to be a prior (like in \cite{theo_bruins}), while $b$ and $h$ are posteriors. In figure \ref{fig:sb}, the experimental data from \cite{exp_cluzel} are shown together with the theoretical curves of \cite{theo_bruins} and the present model. The two models give an equally good description of the experimental data. The force-extension relation of the present model is obtained with a constant temperature $T=300$K and with following values of the model parameters $b=2.9$nm, $a=0.81b$, $h=-20.5b$pN. The typical length of one basepair in B-DNA is $0.34$nm. As a consequence the contour length of B-DNA is equal to $L_B=0.34N_{bp}$nm. Together with (\ref{connection_dna_walk}), one obtains $N_{bp}/N_s\approx4$. So the number of base pairs in one segment is of the order of $4$. In \cite{theo_bruins} the values $1$ and $10$ are used for the number of base pairs in one segment. This means that the two theoretical models ignore interactions that make the DNA molecule stiffer. This is not unexpected, because both models use very simple Hamiltonians. The good representation of the experimental data shows that the theoretical models are able to catch the essence of the overstretching transition when effective parameters are introduced. These effective parameters, like the number of base pairs in one segment, are then used to artificially increase the stiffness of the molecule.

\begin{figure}
\begin{center}
\parbox{0.49\textwidth}{\includegraphics[width=0.48\textwidth]{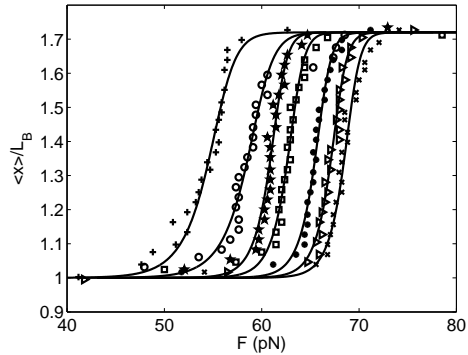}}
\caption{\label{fig:bloom1}This figure shows experimental \cite{exp_bloom} and theoretical force-extension relations obtained for different salt concentrations. The values of the salt concentration (in mM) of the experimental curves are  (left) 10;25;50;100;250;500;1.000 (right). The values of the posteriors of the theoretical curves can be found in table \ref{tab:salt}.} 
\end{center}
\end{figure}

\begin{figure}
\begin{center}
\parbox{0.49\textwidth}{\includegraphics[width=0.48\textwidth]{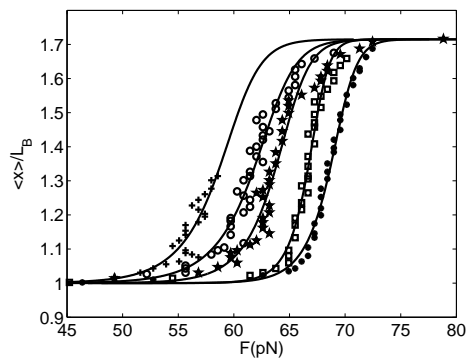}}
\caption{\label{fig:bloom2}This figure shows experimental \cite{exp_bloom2} and theoretical force-extension relations obtained at different temperature. The values of the temperatures (in K) of the experimental curves are (left) 313;308;304;294;284 (right). The values of the posteriors of the theoretical curves can be found in table \ref{tab:temp}. } 
\end{center}
\end{figure}

The two theoretical models contain several adjustable parameters. So a good representation of one dataset is not really surprising. Therefore we test our model further by studying the salt and temperature dependence of the overstretching transition. In figures \ref{fig:bloom1} and \ref{fig:bloom2} the results of our theoretical model are shown together with the experimental data of \cite{exp_bloom} and \cite{exp_bloom2} respectively. In \cite{exp_bloom} the salt dependence of the transition is studied while in \cite{exp_bloom2} the temperature dependence is studied. The value of the prior $a/b$ is equal to $0.86$ for all the theoretical curves. The values of the posteriors are obtained by a least squares analysis. For every curve $\chi^2$ is defined by
\begin{eqnarray}
\chi^2&=&\sum_{i=1}^m\frac{\left(\langle x\rangle_{i,\textrm{theory}}-x_{i,\textrm{experiment}}\right)^2}{\langle x\rangle_{i,\textrm{theory}}},
\end{eqnarray}
with $m$ the number of datapoints of the curve. The values of the posteriors $h/b$ and $h\beta$ are fixed by minimising this expression of $\chi^2$. The result can be found in tables \ref{tab:salt} and \ref{tab:temp}. The sensitivity of the least square analysis is tested by the following procedure. We keep the value of one of the fit parameter fixed (at the value that minimises $\chi^2$) while varying the other. It turns out that the value of $\chi^2$ is doubled after a variation of $10\%$ in the parameter $h\beta$ and after a variation of $0.5\%$ in the parameter $h/b$. This shows that the least squares analysis is more sensitive to variations in $h/b$ than to variations in $h\beta$. Important to mention is that we used the exact formulas of appendix \ref{app} for the least squares analysis, although the approximations of appendix \ref{app2} are very reliable for the values of the parameters used in figures \ref{fig:bloom1} and \ref{fig:bloom2}. This can be seen in figure \ref{fig:approx} where we included a curve with the same values of the adjustable parameters as in the far left curve of figure \ref{fig:bloom1}. 

The theoretical curves are only shown from $\langle x\rangle/L_B\approx1$ to  $\langle x\rangle/L_B\approx1.7$, because the salt dependence (and temperature dependence) at low forces is not well described by the present model. The model contains three parameters $a$, $b$ and $h$. Roughly these parameters determine the factor ($L_S/L_B$) by which the contour length is increased, the value of the overstretching force and the steepness of the transition. When the concentration of salt in the solvent increases, DNA becomes more stable. As a consequence the value of the overstretching force and the steepness of the transition increase, while $L_S/L_B$ remains the same. This can be captured by the present model and results in a decrease of value of the fit parameters $h\beta$ and $h/b$ (see table \ref{tab:salt}). A closer look at the experimental data of \cite{exp_bloom} shows that the force-extension behaviour at low forces depends less on the salt concentration than at intermediate and high forces. This cannot be captured by the present model because the low and high force behaviour cannot be changed independently. In \cite{exp_bloom}, only the influence of Na$^+$ on the overstretching transition is studied. The dependence of the transition on multivalent cations like Mg$^{2+}$ is studied in \cite{multi}. A clear change in the low force behaviour is observed when Mg$^{2+}$ is added to the solution. Our model contains less adjustable parameters in comparison with the other theoretical models studied in the literature. This leaves room for extending our model with one extra parameter which can capture the low force behaviour for different solvent conditions.

\begin{figure}
\begin{center}
\parbox{0.49\textwidth}{\includegraphics[width=0.48\textwidth]{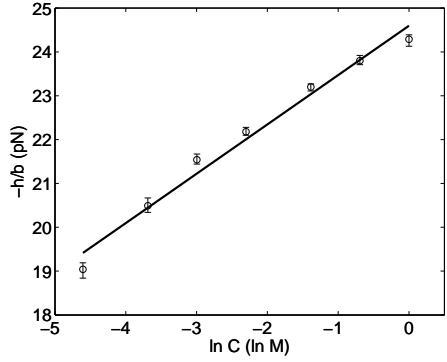}}
\caption{\label{fig:lin}The value of the posterior $h/b$ as a function of the logarithm of the salt concentration.} 
\end{center}
\end{figure}

Figure \ref{fig:lin} shows the value of the parameter $h/b$ as a function of the logarithm of the salt concentration. The error bars in this figure show when the value of $\chi^2$ is doubled by variation of $h/b$ while keeping $h\beta$ fixed. The value of the parameter $h/b$ is approximately a linear function of the logarithm of the salt concentration.

As already mentioned in the introduction, one can assume that force-exten\-sion relations are measured under equilibrium conditions when no hysteresis is observed. In \cite{exp_bloom} one obtains force-extension relations for overstretched DNA at different salt concentrations. The authors mention that they always observe hysteresis in their experiments at low salt concentrations ($<250$mM). In \cite{exp_bloom2} one obtains force-extension relations for overstretched DNA at different temperatures. The authors mention that the stretch and release curves coincide at room temperature, but that the difference between these curves is already $18$pN at $308$K. We conclude that one can assume that $3$ curves of figure \ref{fig:bloom1} ($250$mM, $500$mM and $1.000$mM) and $2$ curves of figure \ref{fig:bloom2} ($294$K and $284$K) are obtained under equilibrium conditions. The assumption of equilibrium is less reliable for the other curves of these figures. It is also important to mention that all curves shown in figures \ref{fig:bloom1} and \ref{fig:bloom2} are obtained during the stretch cycle of the experiment \cite{exp_bloom,exp_bloom2}.
\begin{table}
\begin{center}
\begin{tabular}{|c|ccccccc|}
\hline
$C$ (mM)&10&25&50&100&250&500&1.000
\\\hline
$-h\beta$&24.67&27.54&44.75&46.21&49.44&51.81&52.95
\\
$-h/b$ (pN)&19.04&20.49&21.54&22.18&23.21&23.80&24.29
\\\hline
\end{tabular}
\caption{\label{tab:salt}The values of the posteriors of the theoretical curves of figure \ref{fig:bloom1}.}
\end{center}
\end{table}
\begin{table}
\begin{center}
\begin{tabular}{|c|ccccc|}
\hline
$T$ (K)&313&308&304&294&284
\\\hline
$-h\beta$&20.15&21.01&26.65&45.45&37.01
\\
$-h/b$ (pN)&20.22&21.33&22.16&23.44&24.05
\\\hline
\end{tabular}
\caption{\label{tab:temp}The values of the posteriors of the theoretical curves of figure \ref{fig:bloom2}.}
\end{center}
\end{table}

\section{Titin}\label{appl_tit}
Titin is a long protein that contains approximately $30.000$ amino acids. In the literature, the dynamic force-extension relation of this protein is extensively studied \cite{tit_rev}. In \cite{tsk}, the static (equilibrium) force-extension relation of titin is also measured. 

The protein can be divided into two parts, the I-band and the A-band. At relatively low forces, the major contribution to the elasticity of titin comes from the I-band. Only at very large forces, the A-band becomes important. For that reason we consider only the I-band in this article. The I-band consists of a PEVK region (that contains $1.000-2.200$ amino acids) that is flanked by $70-90$ immunoglobulin domains \cite{tsk}. These immunoglobulin domains are folded parts of the protein containing approximately $100$ amino acids. Without applied force, the chain of amino acids of the PEVK region and the chain of immunoglobulin domains behave like random coils. When one applies a small force, the PEVK region immediately stretches out to almost its complete contour length. Meanwhile, the immunoglobulin domains line up (without unfolding) in the direction of the applied force. At high forces, the immunoglobulin domains will unfold one by one to further increase the contour length of the polymer.

\begin{figure}
\begin{center}
\parbox{0.49\textwidth}{\includegraphics[width=0.48\textwidth]{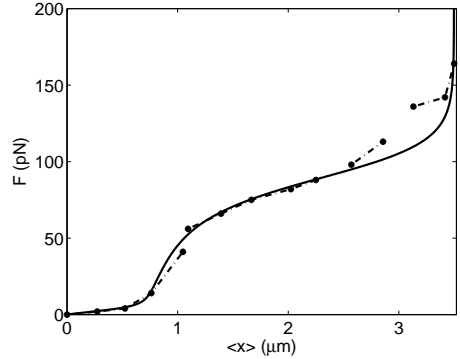}}
\caption{\label{fig:titin}This figure shows the experimental \cite{tsk} (dotted lines) force-extension relation of the protein titin together with the theoretical results (solid line) of the present model. The values of the adjustable parameters are $a_p=b_p=0.35$nm, $a_i=35$nm, $b_i/2=5$nm, $n_p=1995$, $n_i=80$, $h_i/b_i=-180$pN, $h_p/h_i=-3.2$ and $\beta h_i=-0.42$.} 
\end{center}
\end{figure}

We consider the PEVK region and the immunoglobulin domains as two completely independent parts of the protein. The average extension of the complete protein $\langle x\rangle$ is then simply the sum of the average extension of the immunoglobulin domains $\langle x_i\rangle$ and the average extension of the PEVK region $\langle x_p\rangle$
\begin{eqnarray}
\langle x\rangle&=&\langle x_i\rangle+\langle x_p\rangle.
\end{eqnarray}
The indices $i$ and $p$ are used for the variables of the immunoglobulin domains and the PEVK region respectively. A similar formula for the entropy of the complete system holds $S=S_i+S_p$. The Hamiltonian of the complete system is then $H=h_iK_i+h_pK_p$. The sign of $h_i$ is negative because a folded immunoglobulin domain should be energetically favourable. The sign of $h_p$ is positive because the amino acids of the PEVK region behave like a random coil. The free energy of the complete system is
\begin{eqnarray}
G&=&\min_{\epsilon_i,\mu_i,\alpha_i,\gamma_i,\theta_i,\epsilon_p,\mu_p,\alpha_p,\gamma_p,\theta_p}\left\{H-F\langle
x\rangle-\beta S\right\}
\cr
&=&\min_{\epsilon_i,\mu_i,\alpha_i,\gamma_i,\theta_i}\left\{h_i\langle K_i\rangle-F\langle
x_i\rangle-\beta S_i\right\}
\cr
&&+\min_{\epsilon_p,\mu_p,\alpha_p,\gamma_p,\theta_p}\left\{h_p\langle K_p\rangle-F\langle
x_p\rangle-\beta S_p\right\}.
\end{eqnarray}
The results of these two minimalisations can be calculated and are given by (\ref{minim}). Figure \ref{fig:titin} shows the experimental data of \cite{tsk} together with the theoretical results of the present model. The model contains several adjustable parameters. Following parameters can be considered as priors $a_p=b_p=0.35$nm (typical length of one amino acid), $a_i=35$nm (typical length of an unfolded immunoglobulin domain containing $100$ amino acids). The length of a folded domain ($b_i/2$) is less known. Figure \ref{fig:titin} is obtained with the choice $b_i/2=5$nm. This value is well within the range of experimental observations \cite{tsk}. The values of the remaining parameters are $n_p=1995$, $n_i=80$, $h_i/b_i=-180$pN, $h_p/h_i=-3.2$ and $\beta h_i=-0.42$. Note that the values of $n_p$ and $n_i$ are in the range of experimental observations \cite{tsk}. With so many adjustable parameters it is no surprise that figure \ref{fig:titin} shows a very good agreement between the theoretical and experimental results at low and intermediate forces. To the best of our knowledge, a comprehensive experimental study of the static force-extension relation of titin is not yet available in the literature. So an in depth comparison between theoretical and experimental results is not yet possible.

\section{Discussion}
To summarise, we study the three-dimensional persistent random walk with drift. We obtain analytical formulas for the average of the macroscopic variables of interest as a function of the model parameters. Finally, the resulting thermodynamic model is used to study polymer stretching experiments. We derive force-extension relations for two different polymers, overstretched double-stranded DNA and the protein titin. The molecules of double-stranded DNA and of the protein titin are non-oriented. Therefore, each walk and its mirror images must have the same probabilities. This is indeed the case in absence of an external force. However, in the presence of an applied force this is only the case when the distance between the endpoints in the direction of the force does not change under mirroring. We imposed this symmetry in the definition of the transition probabilities (\ref{syymmm}). The substitution $y+\leftrightarrow y-$ has no effect on the transition probabilities, contrary to the substitution $x+\leftrightarrow x-$.

Note that we did not assume the Boltzmann-Gibbs form for the equilibrium probability distribution. Rather, following \cite{erik_resi}, we define the temperature by calculating the Legendre transform of the entropy. The results of \cite{erik_resi} and the present paper show that it is possible to construct a thermodynamic model without the assumption of the Boltzmann-Gibbs distribution. Moreover, we believe that the results of the present paper prove that our approach is more than a mathematical exercise and can be used to study real physical systems. Figures \ref{fig:exte} and \ref{fig:phase} show that the present model contains three separate phases, the random phase (high temperature), the stretched phase (high force) and the compact phase (low temperature and force). In principle a true phase transition can occur in our finite model because we did not assume the Boltzmann-Gibbs form for the equilibrium distribution. To be sure that only gradual transitions show up in our model, we checked that the free energy is a convex function of the control parameters. The transition between the compact and the stretched phase is a gas-liquid-like transition. The boundary line between these two phases ends in an approximate triple point at which the peak in $\partial\langle x\rangle/\partial F$ disappears. 

The applicability of our thermodynamic model is limited to experiments that are performed under equilibrium conditions. Force-extension relations are measured under these circumstances when no hysteresis is observed. The overstretching transition of DNA under equilibrium conditions has been studied more extensively than the force-extension relation of titin. For this reason, the comparison between experimental and theoretical results is more comprehensive for the former than for the latter in the present paper. Figure \ref{fig:titin} shows that our model gives at least a qualitative representation of the experimental static force-extension relation of the protein titin \cite{tsk}. Figures \ref{fig:bloom1} and \ref{fig:bloom2} show that our model gives a consistent description of the overstretching transition of DNA at intermediate and high forces. This means that the force-extension relations obtained with our theoretical model fit well to the experimental curves for varying environmental conditions (different salt concentrations \cite{exp_bloom} and temperatures \cite{exp_bloom2}). In the literature different analytically solvable models are introduced to study the overstretching transition. The model \cite{theo_punk} proposed by Punkkinen et al.~is the only one that is tested against different datasets, obtained under varying salt concentrations. It gives an adequate description over the whole range of the force-extension curve and not only in the transition region. An important difference between our model and the model of Punkkinen et al.~is that the latter model contains more adjustable parameters than our model. This leaves room to extend our model with some extra parameter in order to correct the behaviour at small forces.

In figure \ref{fig:phase} the positive slope of the boundary line, between the compact and the stretched phase of the random walk, is clearly visible. It is an open question whether this so called reentrant behaviour can be observed in any of the experiments that we discussed in the present paper. In \cite{klim}, phase diagrams of the force-induced unfolding of single-domain proteins (e.g.~immunoglobulin domains of titin) are obtained by numerical simulations. These phase diagrams show also reentrant behaviour. As already pointed out in \cite{klim}, the interaction between the solvent and the biopolymer depends on the temperature. As a consequence, the phase diagrams of the present paper and \cite{klim} are only reliable for the study of the stretching experiments on biopolymers for small temperature variations. In the present model, the value of the parameter $h$ is determined by the interaction between the solvent and the biopolymer. This parameter is an effective parameter. Therefore its value can still depend on the temperature (see figure \ref{fig:bloom2} and table \ref{tab:temp}).

In principle one can nowadays perform two types of stretching experiments in two different ensembles. In one type of experiment, the two ends of the molecule are held at fixed positions and the fluctuating force is measured (the fixed-stretch ensemble). In the other type of experiment, the force applied to the endpoints of the molecule is kept constant and the fluctuating extension is measured  (the fixed-force ensemble). Theoretical calculations \cite{erik_aone,bust_fin} show that the force-extension relations obtained in the two ensembles do not coincide in general, although the differences are small and disappear in the long chain limit. The results of the present paper are limited to the fixed-force ensemble, although the discussed experiments are performed in the fixed-stretch ensemble. In \cite{erik_aone,gar}, it is shown how the results of the fixed-force ensemble can be extended to the fixed-stretch ensemble for the one-dimensional persistent random walk with drift. Similar calculations for the three-dimensional random walk are not yet possible because an explicit expression for the joint probability distribution $p_n(x,k)$  is still lacking \cite{erik_atwo,erik_aone}. This is the probability to end in position $x$ with $k$ changes of direction after $n$ steps. Based on the results of the one-dimensional walk \cite{erik_aone} we assume that one can ignore the differences between the two ensembles for the long polymers which are studied in the present paper.

Our results prove that the persistent random walk with drift is an interesting model. Two possible extensions are already suggested above. One can try to improve the applicability of the random walk to the overstretching transition of DNA by introducing one more parameter in the model. One can also try to find an explicit expression for the joint probability distribution $p_n(x,k)$ along the lines of \cite{erik_atwo}. This will not only allow to calculate the force-extension relation in the fixed-stretch ensemble \cite{erik_aone}, but also to quantify the deviations from the Boltzmann-Gibbs distribution \cite{erik_atwo}. Finally, one can also try to extend the applicability of the present model to experiments that are performed under non-equilibrium conditions. To this purpose one can assume, along the lines of superstatistics \cite{beck}, that the model parameters are stochastic variables which have some probability distribution. The major problem is then to determine this time-dependent probability distribution.

\appendix
\section{}\label{app}
Introduce following shorthand notation
\begin{eqnarray}
T_1&:=&\exp\left(\beta\left[-h+F\left(b-a\right)\right]\right)
\cr
T_2&:=&\exp\left(2\beta Fb\right)
\cr 
T_3&:=&\exp\left(2\beta Fa\right)
\cr
T_4&:=&\exp\left(h\beta\right).
\end{eqnarray}
One can calculate explicit expressions for $\gamma$, $\theta$, $\mu$ and $\alpha$ as a function of $\epsilon$ only
\begin{eqnarray}\label{eqqq1}
\mu&=&\frac14\left(1-\frac{1-4\epsilon}{T_3}\right)
\cr
\alpha&=&\frac{T_1}{T_4\sqrt{T_3}}\frac{(1-4\epsilon)^2}{\epsilon}
\cr 
\gamma&=&1-\alpha-\frac{2+T_4}{T_4\sqrt{T_3}}(1-4\epsilon)
\cr
\theta&=&\frac{1}{T_4\sqrt{T_3}}(1-4\epsilon).
\end{eqnarray}
Using these expressions, one is left with following equation for $\epsilon$
\begin{eqnarray}\label{derde}
\epsilon(1-4\epsilon)^2&=&\frac14\left(1-\frac{1-4\epsilon}{T_3}\right)\frac{T_2}{T_1}\Big[-T_1(1-4\epsilon)^2
\cr
&&T_4\sqrt{T_3}\epsilon-(2+T_4)(1-4\epsilon)\epsilon\Big].
\end{eqnarray}
This is a cubic equation that can be solved in closed form. It was checked numerically that is has always a unique physical solution.

\section{}\label{app2}
Assume $h<0$, $a\leq b<2a$ and $\beta\rightarrow\infty$. Then, the cubic equation in $\epsilon$ (\ref{derde}) and the formula for the average endposition (\ref{x}) can be approximated in the transition region ($b/2<\langle x\rangle/n<a$) by following expressions
\begin{eqnarray}\label{bena}
\epsilon=\frac{T_1}{T_4\sqrt{T_3}}(1-4\epsilon)^2&\textrm{and}&\frac{\langle x\rangle}n=\frac{a+4\epsilon(b-a)}{1+4\epsilon}.
\end{eqnarray}
The first equation is now a quadratic equation in $\epsilon$. It has only one physical solution ($\epsilon\in[0..1/4]$). Inserting this solution in the second equality of (\ref{bena}) results in an expression for $\langle x\rangle$ as a function of $\beta$ and $F$. Then one can calculate the second derivative of the average end-to-end distance with respect to the force at constant temperature.
This second derivative equals zero if the following equation holds
\begin{eqnarray}
F\left(1-2\frac ab\right)\frac bh&=&2-\frac1{h\beta}3\ln2.
\end{eqnarray}
This is a low temperature approximation for the boundary line between the
compact and stretched phases. The factor $3\ln2$ is clearly a contribution due to the
entropy. The expressions (\ref{bena}) can also be used to obtain a formula for the force as a function of $\langle x\rangle$ and $\beta$. The result is
\begin{eqnarray}\label{approx_fe}
-Fl\frac bh&=&2+\frac1{h\beta}\Big[\ln\left(l+\chi\right)+\ln\left(l-\chi\right)-2\ln \chi-4\ln2\Big],
\end{eqnarray}
with $\chi=2\langle x\rangle/nb-1$ and $l=2a/b-1$.

\raggedright

\end{document}